\documentclass
[superscriptaddress,secnumarabic,amssymb,amsmath,nobibnotes,aps,prd,showkeys,showpacs,nofootinbib,onecolumn,notitlepage]{revtex4-1}
\usepackage{setspace}
\usepackage{color}
\usepackage{amsmath}
\usepackage{amsfonts}
\usepackage{verbatim}
\usepackage{amssymb}
\usepackage{graphicx,bm}
\usepackage{graphicx}
\usepackage{graphicx}
\usepackage{epstopdf}
\usepackage{epsf}
\usepackage[caption=false]{subfig}
\definecolor{darkgreen}{rgb}{0,0.35,0}

\providecommand{\U}[1]{\protect\rule{.1in}{.1in}}
\begin{document}

\title{Analytic topologically non-trivial solutions of the (3+1)-dimensional
$U(1)$ gauged Skyrme model and extended duality}
\author{L. Avil\'es}
\email{aviles@cecs.cl}
\affiliation{Departamento de F\'isica, Universidad de Concepci\'on, Casilla 160-C, Concepci\'on, Chile}
\affiliation{Centro de Estudios Cient\'{\i}ficos (CECS), Casilla 1469,
Valdivia, Chile}
\author{F. Canfora}
\email{canfora@cecs.cl}
\affiliation{Centro de Estudios Cient\'{\i}ficos (CECS), Casilla 1469,
Valdivia, Chile}
\author{N. Dimakis}
\email{nsdimakis@gmail.com}
\affiliation{Instituto de Ciencias F\'isicas y Matem\'aticas, Universidad
Austral de Chile, Valdivia, Chile}
\author{D. Hidalgo}
\email{dhidalgo@cecs.cl}
\affiliation{Departamento de F\'isica, Universidad de Concepci\'on, Casilla 160-C, Concepci\'on, Chile}
\affiliation{Centro de Estudios Cient\'{\i}ficos (CECS), Casilla 1469,
Valdivia, Chile}

%\author{L. Avil\'es$^{1,2}$, F. Canfora$^{2}$, N. Dimakis$^{3}$, D. Hidalgo$%
%^{1,2}$ \\
%EndAName
%$^{1}$\textit{Departamento de F\'isica, Universidad de Concepci\'on, Casilla 160-C, Concepci\'on, Chile.}\\
%$^{2}$\textit{Centro de Estudios Cient\'{\i}ficos (CECS), Casilla 1469,
%Valdivia, Chile.}\\
%$^{3}$\textit{Instituto de Ciencias F\'isicas y Matem\'aticas, Universidad
%Austral de Chile, Valdivia, Chile.}\\
%{\small aviles@cecs.cl, canfora@cecs.cl, nsdimakis@gmail.com , dhidalgo@cecs.cl}%
%}

\begin{abstract}
We construct the first analytic examples of topologically non-trivial
solutions of the (3+1)dimensional $U(1)$ gauged Skyrme model within a finite
box in (3+1)-dimensional flat space-time. There are two types of gauged
solitons. The first type corresponds to gauged Skyrmions living within a
finite volume. The second corresponds to gauged time-crystals (smooth
solutions of the $U(1)$ gauged Skyrme model whose periodic time-dependence
is protected by a winding number). The notion of electromagnetic duality
can be extended for these two types of configurations in the sense
that the electric and one of the magnetic components can be interchanged. These analytic solutions show very
explicitly the Callan-Witten mechanism (according to which magnetic
monopoles may ``swallow" part of the topological charge of the Skyrmion)
since the electromagnetic field contribute directly to the conserved
topological charge of the gauged Skyrmions. As it happens in
superconductors, the magnetic field is suppressed in the core of the gauged
Skyrmions. On the other hand, the electric field is strongly suppresed in
the core of gauged time crystals.
\end{abstract}

\maketitle

\section{Introduction}

It is impossible to underestimate the relevance of the Skyrme theory \cite%
{skyrme} in high energy physics (for instance see \cite%
{nuc0,nuc1,nuc2,nuc3,nuc4,nuc5}). Not only such a model corresponds to the
low energy QCD \cite{witten0}, but it also discloses beautifully the very
important role of topology in theoretical physics. In particular, the
solitons (\textit{Skyrmions}) of this theory (made of Bosonic degrees of
freedom) describe Baryons, with the Baryon charge being expressed as a
topological invariant (see \cite{finkrub,manton,skyrev1,witten0,giulini,bala0,ANW,guada} and references
therein).

Such tools are, nowadays, very important also in many other areas of physics
such as semiconductors (see \cite{sklat} and references therein),
Bose-Einstein condensates (see \cite{sklat2} and references therein),
magnetic materials (see \cite{MagSkyrme} and references therein),
gravitational physics (see \cite{con01,con02}, \cite%
{bh01,bh03,bh03b,Ioannidou,bh04,bh05} and references therein) and so on.

It is possible to consider the original Skyrme model as a prototype of
non-integrable systems. Until very recently, basically no analytic solution
with non-trivial topological properties was known. In particular, the lack
of explicit solutions with topological charge on flat space-times made very
difficult the analysis of the corresponding phase diagram. Early important
results (based on the original spherical Skyrmion\footnote{%
It is worth emphasizing that both finite volume effects and isospin
chemical potential are expected to break spherical symmetry.}) analyzing
finite density effects as well as the role of the Isospin chemical potential
can be found in \cite{klebanov,chemical1,chemical2,chemical3,chemical4}.

Due to the importance of the Skyrme model as a low energy limit of QCD it is
a mandatory task to analyze the effects of the coupling of a $U(1)$ gauge
field with the Skyrme theory. The so-called gauged Skyrme model is able to
describe the decay of nuclei due to the coupling with weak interactions.
Such a model also describes electric and magnetic properties of Baryons as
well as it also allows to study the decay of nuclei in the vicinity of a
monopole (classic references are \cite{Witten,witten0,gipson,goldstone,dhoker,rubakov}). The gauged Skyrme model is
expected to have very interesting applications in nuclear and particle
physics, as well as in astrophysics, when the coupling of Baryons with
strong electromagnetic fields cannot be neglected.

On the other hand, the field equations of the $U(1)$ gauged Skyrme model are
even more complicated than the field equations of the original Skyrme model.
There are no analytic topologically non-trivial solutions. At a first
glance, one could guess that the task to construct analytic and
topologically non-trivial configurations of the gauged Skyrme model is
hopeless. This is quite unfortunate as, until now, it has been impossible to
construct analytic solutions of the gauged Skyrme model disclosing
explicitly the Callan-Witten mechanism \cite{Witten} (according to which the
Baryon charge of the gauge Skyrme model can be ``swallowed" by the magnetic
field). Another interesting issue is whether or not Skyrmions tend to
suppress the magnetic field within their cores. More generally, very little
is known about the effects of the interactions between $U(1)$ gauge fields
and Skyrmions. Nice numerical studies of topological configurations in the
gauged Skyrme model can be found in \cite{gaugesky1,gaugesky2} and
references therein.

In fact, recently, in (\cite{canfora2,canfora3,canfora4}, \cite%
{canfora4.5,yang1,canfora6,cantalla4,cantalla5} and references therein) an approach has been introduced in order
to build a more general hedgehog ansatz allowing to depart from spherical
symmetry both in Skyrme and Yang-Mills theories (see \cite{canYM1,canYM2,canYM3} and references therein). Such an approach gave rise
to the first (3+1)-dimensional analytic and topologically non-trivial
solutions of the Skyrme-Einstein system in \cite{canfora6} as well as of the
Skyrme model without spherical symmetry living within a finite box in flat
space-times in \cite{Fab1}. In the last reference, it has been possible to
derive also the critical isospin chemical potential beyond which the
Skyrmion living in the box ceases to exist. Due to the similarity of the
minimal $U(1)$ gauge coupling with the introduction of the Isospin chemical
potential, it is natural to wonder whether the results of \cite{canfora3,canfora6} and \cite{Fab1} can be extended to the $U(1)$ gauged Skyrme
model.

Remarkably, using the above approach, it is possible to construct the first
analytic and topologically non-trivial solutions of the $U(1)$ gauged Skyrme
model. There are two types of gauged solitons. Firstly, gauged Skyrmions
living within a finite volume appear as the natural generalization of the
usual Skyrmions living within a finite volume. Secondly, there are smooth solutions of
the $U(1)$ gauged Skyrme model whose periodic time-dependence is protected
by a topological conservation law. These solitons
manifest very interesting similarities with superconductors as well as with
dual superconductors.

This paper is organized as follows: in the second section, the model and
notations are introduced. In the third section a short review of the
properties of the (3+1)-dimensional Skyrme model at finite volume both
without and with isospin chemical potential is presented (such a review is
very useful to understand the novel results in the following sections). In
the fourth section the gauged solitons are constructed and their main
physical properties are discussed. In the fifth section it is discussed how
electromagnetic duality can be extended to include these gauged solitons. In
the sixth section a physically interesting approximation is discussed in
which the Skyrme field is considered as fixed and the electromagnetic field
is slowly turned on. In section \ref{conclusions}, we draw some concluding
ideas.

\section{The $U(1)$ Gauged Skyrme Model}

\label{model}

We consider the $U(1)$ gauged Skyrme model in four dimensions with global $%
SU(2)$ isospin internal symmetry. The action of the system is
\begin{align}
S& =\int d^{4}x\sqrt{-g}\left[ \frac{K}{2}\left( \frac{1}{2}\mathrm{Tr}%
\left( R^{\mu }R_{\mu }\right) +\frac{\lambda }{16}\mathrm{Tr}\left( G_{\mu
\nu }G^{\mu \nu }\right) \right) -\frac{1}{4}F_{\mu \nu }F^{\mu \nu }\right]
\ ,  \label{sky1} \\
R_{\mu }& =U^{-1}D_{\mu }U\ ,\ \ G_{\mu \nu }=\left[ R_{\mu },R_{\nu }\right]
\ ,\ D_{\mu }=\nabla _{\mu }+A_{\mu }\left[ t_{3},\ .\ \right] \ ,
\label{sky2} \\
U& \in SU(2)\ ,\ \ R_{\mu }=R_{\mu }^{j}t_{j}\ ,\ \ t_{j}=i\sigma _{j}\ ,
\label{sky2.5}
\end{align}%
where $\sqrt{-g}$ is the (square root of minus) the determinant of the
metric, $F_{\mu \nu }=\partial _{\nu }A_{\mu }-\partial _{\mu }A_{\nu }$ is
the electromagnetic field strength, $\nabla _{\mu }$ is the partial
derivative, the positive parameters $K$ and $\lambda $ are fixed
experimentally and $\sigma _{j}$ are the Pauli matrices. In our conventions $%
c=\hbar =\mu _{0}=1$, the space-time signature is $(-,+,+,+)$ and Greek
indices run over space-time. The stress-energy tensor is
\begin{equation}
T_{\mu \nu }=-\frac{K}{2}\mathrm{Tr}\left[ R_{\mu }R_{\nu }-\frac{1}{2}%
g_{\mu \nu }R^{\alpha }R_{\alpha }\right. \,+\left. \frac{\lambda }{4}\left(
g^{\alpha \beta }G_{\mu \alpha }G_{\nu \beta }-\frac{g_{\mu \nu }}{4}%
G_{\sigma \rho }G^{\sigma \rho }\right) \right] +\bar{T}_{\mu \nu } , \notag
\label{timunu1}
\end{equation}%
with
\begin{equation}
\bar{T}_{\mu \nu }=F_{\mu \alpha }F_{\nu }^{\;\alpha }-\frac{1}{4}F_{\alpha
\beta }F^{\alpha \beta }g_{\mu \nu },
\end{equation}%
being the part emanating from the electromagnetic action. The matter field equations are
\begin{equation}
D^{\mu }\left( R_{\mu }+\frac{\lambda }{4}\left[ R^{\nu },G_{\mu \nu }\right]
\right) =0\ ,  \label{nonlinearsigma1}
\end{equation}%
\begin{equation}
\nabla _{\mu }F^{\mu \nu }=J^{\nu }\ ,  \label{maxwellskyrme1}
\end{equation}%
where $J^{\nu }$ is the variation of the Skyrme action (the first two terms
in Eq. (\ref{sky1})) with respect to $A_{\nu }$. A direct computation shows
that%
\begin{equation}
J^{\mu }=\frac{K}{2}Tr\left[ \widehat{O}R^{\mu }+\frac{\lambda }{4}\widehat{O%
}\left[ R_{\nu },G^{\mu \nu }\right] \right] \ ,  \label{current}
\end{equation}%
where%
\begin{equation*}
\widehat{O}=U^{-1}t_{3}U-t_{3}\ .
\end{equation*}%
It is worth to note that when the gauge potential reduces to a constant
along the time-like direction, the field equations (\ref{nonlinearsigma1})
describe the Skyrme model at a finite isospin chemical potential.

Hence, the term \textit{gauged Skyrmions} (or, more generically, \textit{%
gauged topological configurations} of the $U(1)$ gauged Skyrme model in
(3+1) dimensions) refers to smooth regular solutions of the coupled system
in Eqs. (\ref{nonlinearsigma1}) and (\ref{maxwellskyrme1}) possessing a
non-vanishing winding number (defined here below in Eq. (\ref{new4.1})). The
aim of the present work is to construct the first (to the best of authors
knowledge) analytic configurations of this type and to disclose their
intriguing physical properties.

\subsection{Topological charge}

If we adopt the standard parametrization of the $SU(2)$-valued scalar $%
U(x^{\mu }) $
\begin{equation}
U^{\pm 1}(x^{\mu })=Y^{0}(x^{\mu })\mathbb{\mathbf{I}}\pm Y^{i}(x^{\mu
})t_{i}\ ,\ \ \left( Y^{0}\right) ^{2}+Y^{i}Y_{i}=1\,,  \label{standnorm}
\end{equation}%
where $\mathbb{\mathbf{I}}$ is the $2\times 2$ identity and
\begin{align}
Y^{0}& =\cos C\ ,\ Y^{i}=n^{i}\cdot \sin C\ ,  \label{pions1} \\
n^{1}& =\sin F\cos G\ ,\ \ n^{2}=\sin F\sin G\ ,\ \ n^{3}=\cos F\ .
\label{pions2}
\end{align}
In the original Skyrme model, without coupling with the $U(1)$ gauge field,
the Skyrme field possesses a non-trivial conserved topological charge. Such
a charge can be expressed as an integral over a suitable three-dimensional
hypersurface $\Sigma $
\begin{equation}
W=\frac{1}{24\pi ^{2}}\int_{\Sigma }\epsilon ^{ijk}Tr\left( U^{-1}\partial
_{i}U\right) \left( U^{-1}\partial _{j}U\right) \left( U^{-1}\partial
_{k}U\right) =\frac{1}{24\pi ^{2}}\int_{\Sigma }\rho _{B}\ .  \label{new4}
\end{equation}%
A direct computation shows that the charge density is $\rho _{B}=12\sin
^{2}C\sin F\ dC\wedge dF\wedge dG$. Obviously, in order for the topological
charge density to be non-vanishing one has to require $dC\wedge dF\wedge
dG\neq 0$.

The usual situation considered in the literature corresponds to a space-like
$\Sigma $ in which case $W$ is the Baryon charge. On the other hand,
recently \cite{Fab1} it has been proposed to also consider cases in which $%
\Sigma $ is time-like or light-like. If $W\neq 0$ (whether $\Sigma $ is
space-like, time-like or light-like) then one cannot deform continuously the
corresponding ansatz into the trivial vacuum $U=\mathbb{\mathbf{I}}$.
Consequently \cite{Fab1} when $\Sigma $ is time-like and $W\neq 0$ one gets
a Skyrmionic configuration whose time-dependence is topologically protected
as it cannot decay in static solutions. These kind of solitons have been
named topologically protected time crystals in \cite{Fab1}.

\subsubsection{Gauged topological charge}

Obviously, when the coupling to a $U(1)$ gauge field is considered, the
expression in Eq. (\ref{new4}) cannot be correct since it is not gauge
invariant. The simplest solution to replace in Eq. (\ref{new4}) all the
derivatives with covariant derivatives is wrong as well (since it leads to a
gauge invariant expression which is, however, not conserved). The correct
solution has been constructed in \cite{Witten} (see also the pedagogical
analysis in \cite{gaugesky1}): the expression for the gauge invariant and
conserved topological charge reads%
\begin{equation}
\begin{split}
W=& \frac{1}{24\pi ^{2}}\int_{\Sigma }\epsilon ^{ijk}Tr\left\{ \left(
U^{-1}\partial _{i}U\right) \left( U^{-1}\partial _{j}U\right) \left(
U^{-1}\partial _{k}U\right) \right. - \\
& \left. \partial _{i}\left[ 3A_{j}t_{3}\left( U^{-1}\partial _{k}U+\partial
_{k}UU^{-1}\right) \right] \right\} .
\end{split}
\label{new4.1}
\end{equation}%
Hence, the topological charge gets one extra contribution which, at the end,
is responsible for the Callan-Witten effect \cite{Witten}. The computations
below show that such an effect (according to which, roughly speaking, a
magnetic monopole may ``swallow" part of the topological charge) is more
general and, in principle, strong magnetic fields may be able to support it
even without magnetic monopoles.

\section{Review of the Skyrmions at finite volume}

In \cite{Fab1} an extension of the method introduced in \cite{canfora6}
which also works in situations in which the Skyrme model is analyzed within
a finite volume in a flat metric has been constructed. It is based on the
following ansatz
\begin{equation}
G=\frac{\gamma +\phi }{2\,}\ ,\ \ \tan F=\frac{\tan H}{\cos A}\ ,\ \ \tan C=%
\frac{\sqrt{1+\tan ^{2}F}}{\tan A}\ ,  \label{pions2.25}
\end{equation}%
where
\begin{equation}
A=\frac{\gamma -\phi }{2\,}\ ,\ \ H=H\left( r,z\right) \ .  \label{pions2.26}
\end{equation}

It can be verified directly that, the topological density $\rho _{B}$ is
non-vanishing. From the standard parametrization of $SU(2)$ \cite{Shnir}
it follows that
\begin{equation}
0\leq \gamma \leq 4\pi ,\quad 0\leq \phi \leq 2\pi \ ,  \label{domain}
\end{equation}%
while the boundary condition for $H$ will be discussed below.

\subsection{Sine-Gordon and Skyrmions}

\label{sine-gordon}

A quite efficient way to put the Skyrme model within a flat region of finite
volume is to introduce the following metric
\begin{equation}
ds^{2}=-dz^{2}+l ^{2}\left( dr^{2}+d\gamma ^{2}+d\phi ^{2}\right) \ ,
\label{Minkowski}
\end{equation}%
(here $z$ is the time variable). The size of the volume of this region is of
order $l ^{3}$\ (the parameter $l $ has dimension of length). On the other
hand $r$, $\gamma $ and $\phi $ are angular coordinates (so that they are
adimensional); the domain of $\gamma $ and $\phi $ is given by \eqref{domain}%
, while for $r$ we choose the finite interval $0\leq r\leq 2\pi $.

In the case in which $A_{\mu }=0$, the gauged Skyrme model reduces to the
original Skyrme configuration and the corresponding field equations (\ref%
{nonlinearsigma1}) can be simplified (without loosing the topological
charge) using the ansatz in Eqs. (\ref{pions1}), (\ref{pions2}), (\ref%
{pions2.25}) and (\ref{pions2.26}) as it has been shown in \cite{Fab1}.
Indeed, one gets\textit{\ just one scalar differential equation for the
profile} $H$
\begin{equation}
\Box H-\frac{\lambda }{8\,l ^{2}(\lambda +2l ^{2})}\sin \left( 4H\right) =0\
,  \label{sineG}
\end{equation}%
where $\Box $\ is the two-dimensional D' Alambert operator.

When $A_{\mu }=0$, the topological Baryon charge $B$ and charge density $%
\rho _{B}$ become respectively
\begin{equation}
B=\frac{1}{24\pi ^{2}}\int_{t=const}\rho _{B}\,,\ \rho _{B}=3\sin
(2H)dHd\gamma d\phi \ .  \label{td1}
\end{equation}

If we replace the topologically non-trivial ansatz in Eqs. (\ref{pions1}), (%
\ref{pions2}), (\ref{pions2.25}) and (\ref{pions2.26}) into the original
action (\ref{sky1}) we obtain an effective action given by
\begin{equation}
\mathcal{L}(H)=16\,l ^{2}(\lambda +2l ^{2})\nabla _{\mu }H\nabla ^{\mu
}H-\lambda \cos (4H),  \label{LagSG}
\end{equation}%
which reproduces equation of motion (\ref{sineG}). The boundary conditions
for the function $H$ are%
\begin{equation}
H(0)=0\ ,\ H(2\pi )=\pm \frac{\pi }{2}\ ,  \label{bc1}
\end{equation}%
which corresponds to $B=\pm 1$ and
\begin{equation}
H(0)-H(2\pi )=0\ ,  \label{bc1.1}
\end{equation}%
which leads to $B=0$. The sector $B=0$ is relevant in the construction of
Skyrmion anti-Skyrmion bound states.

Hence, the original (3+1)-dimensional Skyrme field equations, energy density
and effective action in a topologically non-trivial sector (as $\rho
_{B}\neq 0$) can be reduced to the corresponding quantities of the
(1+1)-dimensional sine-Gordon model (a well known example of integrable
models, see \cite{integrable3}). Following \cite{Fab1}, this allows to
construct Skyrmions as well as Skyrmions anti-Skyrmions bound states%
\footnote{%
Indeed, a quite remarkable prediction of the Skyrme model at finite volume
(as discussed in \cite{Fab1}) is that the model possesses around $8\pi
/\beta ^{2}-1$ Skyrmions anti-Skyrmions bound states where $\beta $ is in
Eq. (\ref{SG2}). When the size of the box is large compared with $fm$ one
gets that the number of these bound states is between 5 and 6 (in good
agreement with the number of Baryons anti-Baryons resonances appearing in
particles physics).}. The effective coupling sine-Gordon action and coupling
constants (following \cite{coleman}) read%
\begin{equation}
\mathcal{L}(\Phi )=-\frac{1}{2}\nabla ^{\mu }\Phi \nabla _{\mu }\Phi +\frac{%
\alpha }{\beta ^{2}}\left( \cos \left( \beta \Phi \right) -1\right) \ ,
\label{SG1}
\end{equation}%
\begin{equation}
\alpha =\frac{\lambda }{2l ^{2}\left( \lambda +2l ^{2}\right) },\quad \beta =%
\frac{4l }{\sqrt{\lambda +2l ^{2}}}\ .  \label{SG2}
\end{equation}%
Therefore the Skyrme model within the finite volume defined above always
satisfies the Coleman bound $\beta ^{2}<8\pi $.

It is worth emphasizing that Skyrme and Perring \cite{SkPer} used
sine-Gordon in ($1+1$)-dimensions as a \textquotedblleft toy model" for the ($%
3+1$)-dimensional Skyrme model. The analogies between (a simplified version
of) the Skyrme model and the sine-Gordon model have also been emphasized in
\cite{sinegordan2} and references therein. The very surprising feature of
the results in \cite{Fab1} is that there is a nontrivial topological sector
of the full ($3+1$)-dimensional Skyrme model in which it is \textit{exactly
equivalent to the sine-Gordon model} in ($1+1$)-dimensions\footnote{%
The semi-classical quantization in the present sector of the Skyrme model
can be analyzed following \cite{ANW} \cite{bala0} \cite{palais}: since
\textit{principle of symmetric criticality} applies (see \cite{Fab1}).}.

\subsubsection{An interesting function}

\label{secintfunc}

As it is well known, in the usual case the energy of the spherical Skyrmion
(found numerically by Skyrme himself) exceeds the bound in terms of the
Baryon charge by 23\%. One can ask the similar question on the Skyrmions
living at finite volume constructed in \cite{Fab1}. Only in this subsection,
we will adopt the convention that $K=2$ and $\lambda =1$ (see page 25 of
\cite{skyrev1}) according to which lengths are measured in $fm$ and energy
in $GeV$. Let us consider the function
\begin{equation}\label{bound}
\Delta =E-12\pi ^{2}\left\vert B\right\vert =E-12\pi ^{2},
\end{equation}
where $E$ is the energy of the (anti)Skyrmion constructed in \cite{Fab1} and
$B$ is its baryon charge. The nice pedagogical review \cite{Zahed} clarifies
that Eq. (\ref{bound}) corresponds to the right hand side of the Skyrme BPS
bound in terms of the Baryon charge. The function $\Delta $ (once $K$ and $%
\lambda $ have been fixed) is a function of the size of the box $l$ in
which these configurations live. We have to note that in the relevant equation in \cite{Fab1} appears a $\sqrt{2}$ multiplying the term that is substracted from the energy, which is a typo. One can see that for high densities (when $l$ is around 10 $fm$ or less) the
Skyrmions constructed exceed the topological bound by $\sim 53\%$
and when the size is large, their energy increases very rapidly.

\subsubsection{Inclusion of chemical potential and infinite volume limit}

As it is well known, the presence of the isospin chemical potential is
encoded in the following covariant derivative (see \cite{chemical1} \cite%
{chemical2} \cite{chemical3} \cite{chemical4})
\begin{equation}
D_{\mu }=\nabla _{\mu }+\bar{\mu}[t_{3},\;\;]\delta _{\mu 0}.
\label{newcovdiv}
\end{equation}%
This has a correspondence to a special case of a coupling with an
electromagnetic field with a potential of the form $A_{\mu }=\bar{\mu}\delta
_{\mu 0}$. As it has been shown in \cite{Fab1} that for static
configurations $H(r)$ \textit{the full Skyrme field equations with isospin
chemical potential reduce to the following ODE for}
\begin{equation}
\begin{split}
& \left( \lambda +2\,l^{2}-8\,\lambda \,l^{2}\bar{\mu}^{2}\sin
^{2}(H(r))\right) H^{\prime \prime }(r)-4\lambda l^{2}\bar{\mu}^{2}\sin
(2H(r))H^{\prime 2} \\
& +\lambda \left( \,\bar{\mu}^{2}l^{2}-\frac{1}{8}\right) \sin (4H(r))+4\,%
\bar{\mu}^{2}l^{4}\sin (2H(r))=0\ ,
\end{split}
\label{chempotODE}
\end{equation}%
which can be further reduced to%
\begin{equation}
Y\left( H\right) \frac{\left( H^{\ \prime }\right) ^{2}}{2}+V\left( H\right)
=E_{0}\ ,\ \   \label{chempotODE1}
\end{equation}%
where
\begin{eqnarray*}
Y\left( H\right)  &=&\lambda +2\,l^{2}-8\,\lambda \,l^{2}\bar{\mu}^{2}\sin
^{2}(H),\  \\
V\left( H\right)  &=&-\frac{\lambda }{4}\left( \,\bar{\mu}^{2}l^{2}-\frac{1}{%
8}\right) \cos (4H)-2\,\bar{\mu}^{2}l^{4}\cos (2H)\ .
\end{eqnarray*}%
In order to determine the integration constant $E_{0}$ one has to require
the relation here below%
\begin{equation}
\int_{0}^{\pi /2}\frac{\left[ Y\left( H\right) \right] ^{1/2}}{\left[
E_{0}-V\left( H\right) \right] ^{1/2}}dH=\sqrt{2}2\pi \ .  \label{phybc}
\end{equation}%
Consequently, the critical isospin chemical potential $\overline{\mu }_{c}$
can be defined as the value of $\mu $ beyond which Eq. (\ref{phybc}) cannot
be satisfied anymore:%
\begin{equation}
\left( \overline{\mu }_{c}\right) ^{2}=\frac{\lambda +2\,l^{2}}{8\,\lambda
\,l^{2}}\ .  \label{critmu}
\end{equation}%
It is also easy to see that, before reaching the critical value defined
above, the presence of the chemical potential suppresses the energy-peak of
the Skyrmion making it flatter. This comment will be useful to provide with
a simple interpretation of the physical effects of the $U(1)$ gauge coupling.

It is worth emphasizing that if one considers the infinite volume limit of the above
expression for the critical Isospin chemical potential one gets a value which is consistent
with the value computed in the literature (see references \cite{chemical3} and \cite{chemical4}) in the standard infinite volume case. We hope to
come back on the relations between our finite-volume results and the infinite volume limit
in a future publication.

\subsection{Time crystals}

\label{time crystal}

In \cite{timec1,timec2,timec3}, Wilczek and Shapere made the
following deep observation. One can construct simple models in which it is%
\textit{\ possible to break spontaneously time translation symmetry.}

Although it is well-known that no-go theorems \cite{timec4,timec5}
ruled out the original proposals, new research fields started trying to
realize in a concrete system the ideas presented in \cite{timec1,timec2} and
\cite{timec3} (a nice review is \cite{timecr}). Many examples have been
found since then in condensed matter physics \cite{timec5.5,timec5.6,timec5.7,timec6,timec7,timec9}. The first example
in nuclear and particles physics has been found in \cite{Fab1} in the Skyrme
model at finite volume.

Namely, the (3+1)-dimensional Skyrme model supports exact time-periodic
configurations which cannot be deformed continuously to the trivial vacuum
as they possess a non-trivial winding number. Consequently, these time
crystals are only allowed to decay into other time-periodic configurations:
hence, the name \textit{topologically protected time crystals}.

Needless to say, there are many time-periodic solutions of the Skyrme model
which cannot be considered time-crystals\footnote{%
As an example, consider the Skyrmion--anti-Skyrmion bound state at finite
volume which corresponds to breather (so that they are time-periodic). In
fact, they are not topologically protected since, if one `pays' the
corresponding binding energies, they decay into the trivial vacuum.}.

Following \cite{Fab1}, a good choice to describe the finite box is the line
element
\begin{equation}
ds^{2}=-d\gamma ^{2}+l^{2}\left( dz^{2}+dr^{2}+d\phi ^{2}\right) \ ,
\label{metric3}
\end{equation}%
where $\gamma $ plays the role of time. We have to make the following
modification to ansatz (\ref{pions2.25}), (\ref{pions2.26})
\begin{equation}
A=\frac{\omega \gamma -\phi }{2}\,,\quad G=\frac{\omega \gamma +\phi }{2}\,,
\label{ansnew}
\end{equation}%
where $0\leq \omega \gamma \leq 4\pi $ and the frequency $\omega $ is
necessary to keep $A$ and $G$ dimensionless. Note that, with the above
ansatz, the Skyrme configuration $U$ is periodic in time.

The profile $H$ depends on two space-like coordinates. In the case in which
the coupling with the $U(1)$ gauge field is neglected, the Skyrme
configurations $U$ defined in Eqs. (\ref{pions1}), (\ref{pions2}), (\ref%
{pions2.25}), (\ref{pions2.26}) and (\ref{ansnew}) are necessarily
time-periodic. The full Skyrme field equations (\ref{nonlinearsigma1})
reduce in this case to
\begin{equation}
\triangle H-\frac{\lambda \omega ^{2}}{4\left( l^{2}(\lambda \omega
^{2}-4)-\lambda \right) }\sin (4H)=0\ ,  \label{thirdeqH}
\end{equation}%
\begin{equation}
\omega ^{2}\neq \omega _{c}^{2}=\frac{\lambda +4l^{2}}{l^{2}\lambda }\ ,
\label{crifre}
\end{equation}%
where $\triangle $\ is the two-dimensional Laplacian in $z$ and $r$. Eq. (%
\ref{thirdeqH}) is the Euclidean sine-Gordon equation. Exact solutions of
Eq. (\ref{thirdeqH})\footnote{%
Previous literature on the analogies between sine-Gordon and Skyme models
can be found in \cite{sin0,sin1,sin2} and references therein. As it has been
emphasized previously, sine-Gordon theory was believed to be just a
\textquotedblleft toy model" for the 3+1 dimensional Skyrme model. In fact,
we proved that in a nontrivial topological sector they exactly coincide.}
can easily be constructed taking, for instance, $H=H\left( r\right) $.

Time crystal configuration can be constructed explicitly considering $%
H=H\left( r\right) $. These configurations have a non-trivial winding
number. The topological density is given by $\rho _{B}=3\sin (2H)dH\wedge
d\left( \omega \gamma \right) \wedge d\phi $, and thus the winding number
can be evaluated to
\begin{equation}
W=\frac{\omega }{8\pi ^{2}}\int_{z=\text{const}}\sin (2H)dHd\gamma d\phi
=\pm 1.
\end{equation}%
Hence, there are smooth time-periodic regular configurations of the Skyrme
model living at finite volume possessing a non-trivial winding number along
a three-dimensional time-like surface. Consequently, these configurations
can only decay into other configurations which are also time-periodic (as
for static configurations the above winding number vanishes). Thus, the time
periodicity of these configurations is topologically protected by their
winding number.

It is worth stressing the following fact: In the well known case of
non-Abelian gauge theories admitting BPS monopoles, the ground state in the
sector with unit non-Abelian magnetic charge is the BPS monopole. Such
configuration cannot be deformed continuosly to the trivial vacuum: in
particular is not invariant under spatial rotations (unless they are
compensated by internal rotations). The famous \textquotedblleft spin from
isospin effect\textquotedblright\ disclosed in the seventies is a
consequence of this lack of invariance. In the topologically protected
time-crystals constructed in \cite{Fab1}, the ground state is time-periodic
and consequently the theorems in \cite{timec4} and \cite{timec5} (which
assume that the ground state is static) do not apply, in complete analogy
with what happens for BPS monopoles.

\subsubsection{The chemical potential}

One can introduce the isospin chemical potential also for these
time-crystals. The analysis in \cite{Fab1} shows that critical chemical
potential $\overline{\mu }^{\ast }$ can be determined by requiring%
\begin{equation}
\lambda +l^{2}\left[ 4\,-\lambda \omega ^{2}+8\,\lambda \,\overline{\mu }%
^{\ast }\left( \omega -2\overline{\mu }^{\ast }\right) \right] \leq 0\ .
\label{critime}
\end{equation}
The latter condition implies
\begin{equation}
  \overline{\mu }^{\ast } \leq \frac{\omega }{4}-\frac{\sqrt{\frac{4 l^2}{\lambda }+1}}{4 l} \quad \text{or} \quad \overline{\mu }^{\ast } \geq \frac{\omega }{4}+\frac{\sqrt{\frac{4 l^2}{\lambda }+1}}{4 l} ,
\end{equation}
which leads us to consider as critical values
\begin{equation}
  \overline{\mu }_{cr}^{\ast } =\frac{\omega }{4}\pm\frac{\sqrt{\frac{4 l^2}{\lambda }+1}}{4 l} .
\end{equation}

\section{Gauged Skyrmions and Time Crystals}

In this section, we extend the Skyrmions and the time crystals constructed
above to the cases in which the minimal coupling with the $U(1)$ gauge field
cannot be neglected: namely, we will construct analytic examples of gauged
Skyrmions as well as gauged time crystals. Then, in the following sections
we analyze the most interesting physical properties of these gauged
configurations.

For what follows we start by considering the following parametrization of the $SU(2)$%
-valued scalar $U$
\begin{equation}
U = e^{t_3 \alpha} e^{t_2 \beta} e^{t_3 \rho},
\end{equation}
where $\alpha$, $\beta$ and $\rho$ are the Euler angles which in a single
covering of space take the values $\alpha \in [0,2\pi]$, $\beta \in [0,\frac{%
\pi}{2}]$ and $\rho \in [0,\pi]$.

\subsection{Gauged Skyrmions}

Like in the case without an electromagnetic field we start by considering
metric \eqref{Minkowski}, where the ordering of the coordinates that we
assume is
\begin{equation}
x^{\mu }=\left( z,r,\gamma ,\phi \right),   \notag
\end{equation}%
and where again we fix the dimension of the spatial box by requiring
\begin{equation}
0\leq r\leq 2\pi \quad 0\leq \gamma \leq 4\pi ,\quad 0\leq \phi \leq 2\pi \ .
\end{equation}%
As before, $l$ represents the size of the box while $r$, $\gamma $ and $\phi
$\ are adimensional angular coordinates and $z$ represents the time
coordinate. It is possible to choose the ansatz for the Skyrme
configuration in the following manner:
\begin{equation}
\alpha =p\frac{\gamma }{2},\,\beta =H(r),\,\rho =q\frac{\phi }{2}\ ,\ \ p,\
q\in
%TCIMACRO{\U{2115} }%
%BeginExpansion
\mathbb{N}
%EndExpansion
\ ,  \label{ans1}
\end{equation}%
where $p$ and $q\ $\ must be integer in order to cover $SU(2)$ an integer
number of times. We restrict ourselves to the study of a static profile $%
H=H(r)$. In this context we assume an electromagnetic potential of the form
\begin{equation}
A_{\mu }=(b_{1}(r),0,b_{2}(r),b_{3}(r))\ .  \label{EMpotans1}
\end{equation}

Under the previous setting, the ensuing Maxwell equations \eqref{maxwellskyrme1}
become
\begin{equation}  \label{Maxgen}
b_I^{\prime \prime }(r) =\frac{K}{2} \left( M_{IJ} b_J(r) + N_I \right), \quad  I,J=1,2,3
\end{equation}
where
\begin{subequations} \label{Mmatrix1}
\begin{align}
& M_{11} = 4 \sin ^2(H) \left(2 \lambda H^{\prime 2}+ \frac{%
\lambda\left(p^2+q^2\right)}{2} \cos ^2(H)+ 2 l^2\right), \\
& M_{23} = M_{32} = -\frac{p q}{2} \lambda \sin^2(2 H), \\
& M_{22} = M_{11} + \frac{p}{q} M_{23}, \\
& M_{33} = M_{11} + \frac{q}{p} M_{23},
\end{align}
the rest of $M_{IJ}$'s zero and
\end{subequations}
\begin{equation} \label{Nvector1}
N = (0,\frac{p}{4} M_{11} - \frac{q^2-p^2}{4 q}M_{23}, -\frac{q}{4}M_{11} -
\frac{q^2-p^2}{4 p}M_{23}).
\end{equation}

Quite remarkably, the hedgehog property is not destroyed by the coupling to
the above $U(1)$ gauge field since \textit{the Skyrme equations lead to a
single equation for the profile} $H(r)$ (see the appendix for more details)
\begin{equation}
\begin{split}
& 4\left( X_{1}\sin ^{2}(H)+\frac{\lambda (p^{2}+q^{2})}{2}+2l^{2}\right)
H^{\prime \prime }+2X_{1}\sin (2H)H^{\prime 2}+4\sin ^{2}(H)X_{1}^{\prime
}H^{\prime } \\
& +\left( 2\lambda \left( pb_{2}+qb_{3}\right) \left( pb_{2}+qb_{3}+\frac{%
p^{2}-q^{2}}{2}\right) -\frac{\lambda p^{2}q^{2}}{2}-\frac{p^{2}+q^{2}}{2}%
X_{1}\right) \sin (4H)-\frac{2l^{2}X_{1}}{\lambda }\sin (2H)=0,
\end{split}
\label{skyrmeeq1}
\end{equation}%
where
\begin{equation}
X_{1}(r)=4\lambda \left(
-2l^{2}b_{1}^{2}+b_{2}(2b_{2}+p)+b_{3}(2b_{3}-q)\right) .  \label{prex1}
\end{equation}

Still, at a first glance, it is a quite hopeless task to find analytic
solutions to the coupled system corresponding to Eqs. (\ref{Maxgen}) and (%
\ref{skyrmeeq1}).

In fact, a closer look at the simpler situation (described in the previous
section) in which one wants to describe the effects of the isospin chemical
potential offers a surprising solution.

\textit{Firstly}, one has to observe that the Skyrme field equations Eqs. (%
\ref{chempotODE}) and (\ref{chempotODE1}) are integrable (as they are
reduced to quadratures).

\textit{Secondly}, one can ask the following question: under which
circumstances Eq. (\ref{skyrmeeq1}) for the Skyrme profile interacting with
the $U(1)$ gauge field becomes as similar as possible to (the much easier)
Eq. (\ref{chempotODE})?

The answer is that this happens in the special cases where
\begin{equation}
X_{1}=-\frac{\lambda (p^{2}+q^{2})}{2}=\text{const.}\ ,  \label{condin1}
\end{equation}%
and
\begin{equation}
b_{2}(r)=-\frac{q}{p}b_{3}(r)-\frac{p^{2}-q^{2}}{4p}\ .  \label{condin2}
\end{equation}%
Thus, when Eqs. (\ref{condin1}) and (\ref{condin2}) are satisfied, Eq. (\ref%
{skyrmeeq1}) for the Skyrme profile interacting with the $U(1)$ gauge field
becomes integrable (as it can be reduced to a quadrature using the same step
to go from Eq. (\ref{chempotODE}) to Eq. (\ref{chempotODE1})).

However, we are not done yet since Eqs. (\ref{condin1}) and (\ref{condin2})
could be incompatible with Maxwell equations Eq. (\ref{Maxgen}). In other
words, it could happen that there is no solution of Eq. (\ref{Maxgen}) in
which Eqs. (\ref{condin1}) and (\ref{condin2}) are satisfied.

In fact, a direct computation shows that if one replaces Eqs. (\ref{condin1}%
) and (\ref{condin2}) into Maxwell equations Eq. (\ref{Maxgen}), then the
full system reduces to the following single scalar equation for $b_{3}(r)$
\begin{equation}
b_{3}^{\prime \prime }+\frac{K}{4}(q-4b_{3})\sin ^{2}(H)\left(
4l^{2}+4\lambda H^{\prime 2}+\lambda \left( p^{2}+q^{2}\right) \cos
^{2}(H)\right) =0  \label{fullm1}
\end{equation}%
and the corresponding equation for the profile $H$ reads
\begin{equation}
\left( \frac{8l^{2}}{p^{2}+q^{2}}+2\lambda \cos ^{2}(H)\right) H^{\prime
\prime }+\sin (2H)\left( l^{2}-\lambda H^{\prime 2}\right) =0\ .
\label{fullsk1}
\end{equation}%
Interestingly enough, the above equation for the profile interacting with a $U(1)$ gauge field is equivalent to the Skyrme field equation with a chemical potential possessing a value $\bar{\mu}_{0}^{2}=\frac{p^{2}q^{2}%
}{4l^{2}(p^{2}+q^{2})}$.

This is a quite remarkable result \textit{since the full coupled Skyrme
Maxwell system} made by Eqs. (\ref{nonlinearsigma1}) and (\ref%
{maxwellskyrme1}) in a topologically non-trivial sector (as it will be shown
below) in the finite box defined in Eq. \eqref{Minkowski} \textit{can be
reduced consistently to a solvable system of two ODEs} (namely, Eqs. (\ref%
{fullm1}) and (\ref{fullsk1})). Hence, gauged Skyrmions can be constructed
explicitly. In the appendix there can be found the details of the derivation of this result.

The recipe is to use the static ansatz in Eq. (\ref{ans1}) for the Skyrme
configuration and the ansatz in Eqs. (\ref{EMpotans1}), (\ref{condin1}) and (%
\ref{condin2}) for the $U(1)$ gauge field. Thus, one can determine the
Skyrme profile $H(r)$ from Eq. (\ref{fullsk1}) and then Eq. (\ref{fullm1})
for the gauge potential $b_{3}(r)$ becomes a simple linear non-homogeneous
equation in which there is an effective potential which depends on $H(r)$.
The other components of the gauge potential are determined solving the
simple algebraic conditions (\ref{condin1}) and (\ref{condin2}). The above
system allows to clearly disclose many features of the $U(1)$ gauged Skyrme
model which are close to superconductivity.

In comparison with the chemical potential \eqref{chempotODE}, where $p=q=1$,
the corresponding value $\bar{\mu}_{0}^{2}=\frac{1}{8l^{2}}$ is lower than
the upper critical bound set as, $\bar{\mu}_{c}^{2}=\frac{1}{8l^{2}}+\frac{1%
}{4\lambda }$ by \eqref{critmu}. It can be shown that the same thing happens
with the introduction of $p$ and $q$ since now - by following the same
procedure - the critical value ends up to be $\bar{\mu}_{c}^{2}=\frac{1}{%
16l^{2}}(p^{2}+q^{2})+\frac{1}{4\lambda }$ and again $\bar{\mu}_{0}\leq \bar{%
\mu}_{c}$ for $\lambda >0$ and any value of $p$, $q$. Thus, the first
physical conclusion can be drawn. Since the equation for the Skyrme profile
coupled with the $U(1)$ gauge field looks like the Skyrme field equations
with isospin chemical potential and we know that a non-vanishing
isospin chemical potential suppresses the Skyrmion (until it reaches the
critical value when the Skyrmion completely disappears), we can conclude
the coupling with the Maxwell field suppresses (but without destroying) the
Skyrmion.

As a consistency check, if one considers $b_{i}\rightarrow 0\Rightarrow
X_{1}\rightarrow 0$, then \eqref{skyrmeeq1} reduces to
\begin{equation}
H^{\prime \prime }(r)-\frac{\lambda p^{2}q^{2}\sin (4H(r))}{4\left(
4l^{2}+\lambda (p^{2}+q^{2})\right) }=0,  \label{skyrmered1}
\end{equation}%
with a conserved quantity of
\begin{equation}
H^{\prime 2}+\frac{\lambda p^{2}q^{2}\cos (4H(r))}{8\left( 4l^{2}+\lambda
(p^{2}+q^{2})\right) }=I_{0}=\text{const.}
\end{equation}%
The general solution of \eqref{skyrmered1} is given in terms of the Jacobi
amplitude function.

\subsubsection{Topological charge}

Here we calculate the temporal component of the baryon density as it is
modified by the Maxwell field following the steps of \cite{Witten}. Due to $%
A_{\mu}$ and the introduction of a covariant derivative there exists is an
additional term in the form of a total divergence and the full density reads
\begin{equation}  \label{tcharge}
\begin{split}
B_0 & = \frac{\epsilon_{0ijk}}{24 \pi^2} \Big[\mathrm{Tr} \left( R_i R_j R_k
\right) - 3 \partial_{i} \left(A_j \mathrm{Tr} \left( \tau_3\left( U^{-1}
\partial_k U + \partial_k U U^{-1} \right) \right) \right) \Big] \\
& = - \frac{p q}{8\pi^2} \sin(2H) H^{\prime }+ \frac{p q}{4\pi^2} \partial_r
\left( \cos^2(H) \left( b_2-b_3\right)\right) .
\end{split}%
\end{equation}

The Baryon number that we get with the help of $B_{0}$ is
\begin{equation}
B=\int B_{0}drd\gamma d\phi =-p\,q\int \sin (2H)dH+2\Big[\cos
^{2}(H(r))\left( q\,b_{2}(r)-p\,b_{3}(r)\right) \Big]_{0}^{2\pi },
\end{equation}%
and it leads to
\begin{subequations}
\label{topchsk}
\begin{align}
B=& -pq-2\left( q\,b_{2}(0)-p\,b_{3}(0)\right),  \\
B=& pq+2\left( q\,b_{2}(2\pi )-p\,b_{3}(2\pi )\right),
\end{align}%
\end{subequations}
depending on the boundary values that we assume: $H(2\pi )=\pi /2$, $H(0)=0$
or $H(2\pi )=0$, $H(0)=\pi /2$ respectively. Clearly, $B$ depends now on the
size of the system through $p$ and $q$, as well as on the boundary values
set for $b_{2}$ and $b_{3}$ which are related to the magnetic components of $%
F_{\mu \nu }$. The solutions we have found with $p=q=1$ for $b_{3}$ (and the
corresponding values of $b_{2}$) of Eq. (\ref{fullm1}) have $%
b_{2}(0)=\,b_{3}(0)$ so that the topological charge reduces to the usual
integer value. However, it is clear that there are much more general
possibilities and one could try to construct configurations in which the
topological charged is \textquotedblleft shared" by the Skyrmion and the
electromagnetic field. We hope to come back on this interesting issue in a
future publication.

It is worth noting that by combining tools developed in the present paper with the
techniques introduced in \cite{cantalla4} one can construct multi-layered configurations of the gauged Skyrme model such that each layer corresponds to the present gauged-Skyrmion configuration with Baryon charge the product $pq$, while the number of layers is related to the number of peaks of the (energy-density associated to the) profile $H(r)$. This observation suggests that the present formalism could be used to describe analytically the regular patterns which are
known to appear in the Skyrme model when fine-density effects are taken into account. We hope to come back on this very interesting issue in a future publication.

\begin{figure}[h]
\centering
\subfloat[][Energy density $T_{00}$ of the Skyrmion.]{\includegraphics[width=.40 \textwidth]{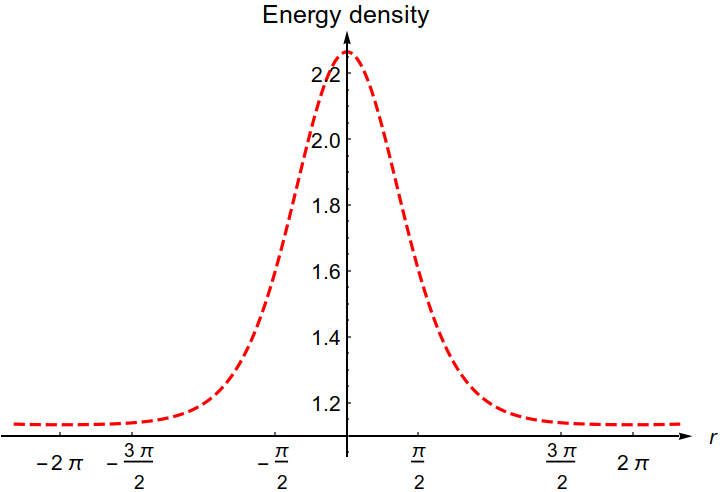}}
\hspace{0mm}
\subfloat[][Skyrme profile.]{
	\includegraphics[width=.40 \textwidth]{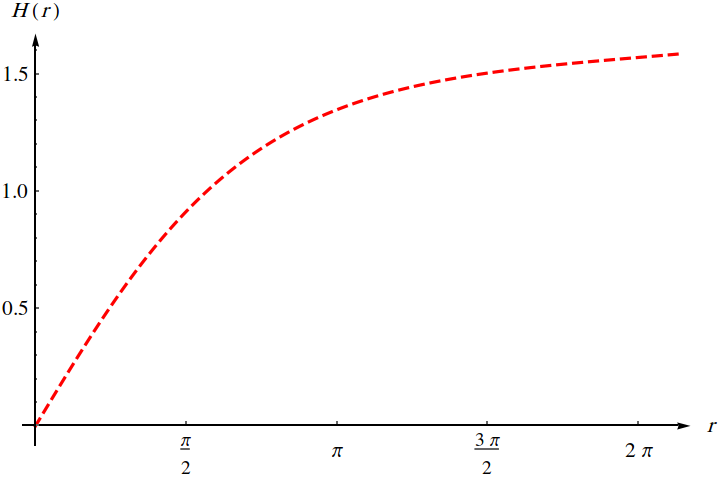}}
\hspace{0mm}
\subfloat[][Gauge potential $A_{\mu}$.]{\includegraphics[width=.40\textwidth]{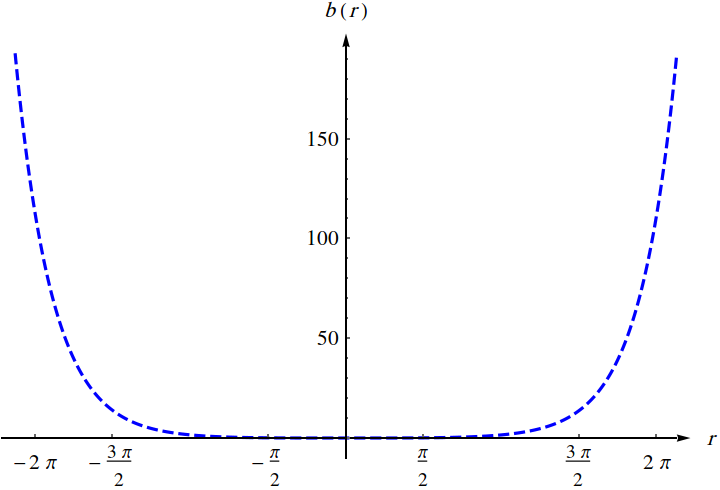}
}
\hspace{0mm}
\subfloat[][Magnetic field $B\equiv b'(r)$.]{\includegraphics[width=.40 \textwidth]{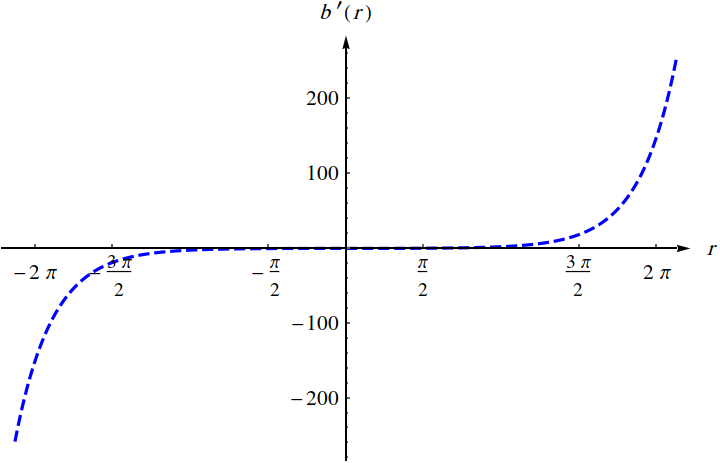} }
\caption{The solutions for the Eqs.\eqref{fullm1} and \eqref{fullsk1} correspond to the values: $\lambda=0.04$, $l=0.47$, $K=1.0$, $p=1.0$, and $q=1.0$. Solving for $b\equiv -b_2=b_3 $. The above plots clearly show the suppression of the magnetic field (which is non-vanishing only in the $\gamma$ and $\phi$ directions) in the core of the Skyrmion.} \label{Fig1}
\end{figure}

\subsection{Gauged time crystal}

As in \cite{Fab1}, a very efficient choice to describe the finite box is the
line element in Eq. (\ref{metric3}) where $\gamma $ plays the role of time.
The Skyrme configuration reads
\begin{equation}  \label{eultc}
\alpha =\frac{\phi }{2},\,\beta =H(r),\,\rho =\frac{\omega \gamma }{2},
\end{equation}
where $\omega $ again is a frequency so that $\rho $ is dimensionless (as it
should be). Once more we assume an electromagnetic potential of the form %
\eqref{EMpotans1}, but now we have to consider that the coordinate ordering
is
\begin{equation}
x^{\mu }=(\gamma ,r,z,\phi ).  \label{cotc}
\end{equation}
The Maxwell equations have the same form as \eqref{Maxgen}. The entries of
matrix $M$ read
\begin{subequations} \label{Mmatrix2}
\begin{align}
& M_{11}= 2 \sin ^2(H(r)) \left(4 \lambda H^{\prime 2}+\lambda \cos ^2(H)+ 4
l^2\right), \\
& M_{13}= -\frac{\lambda \omega}{2} \sin ^2(2 H), \\
& M_{22}= M_{11}+ \frac{l^2 \omega ^2+1}{\omega } M_{13}, \\
& M_{33}= M_{11}+ l^2 \omega M_{13}, \\
& M_{31}=-l^2 M_{13},
\end{align}
while
\end{subequations}
\begin{equation} \label{Nvector2}
N= \left(\frac{1}{4}(M_{13}-\omega M_{11}), 0, \frac{1}{4} \left(\frac{%
\left(2 l^2 \omega ^2+1\right)}{\omega }M_{13}+M_{11}\right)\right) .
\end{equation}

Interestingly enough, also in this case the hedgehog property is not lost.
Namely, the full Skyrme field equations for the time-crystal ansatz defined
above coupled to the $U(1)$ gauge field in Eq. \eqref{EMpotans1} (taking
into account that the coordinates are as in Eq. (\ref{cotc})) reduce to a
single ODE for the profile $H(r)$ (for more details see the appendix)
\begin{equation}  \label{tcprof}
\begin{split}
& 4\left( l^{2}\left( 4-\lambda \omega ^{2}\right) +X_{2}\sin
^{2}(H)+\lambda \right) H^{\prime \prime }+2X_{2}\sin (2H)H^{\prime 2}+4\sin
^{2}(H)X_{2}^{\prime }H^{\prime } \\
& +\left[ \frac{1}{4}\left( l^{2}\omega ^{2}-1\right) X_{2}+\lambda \left(
2l^{2}\omega b_{1}-2b_{3}-1\right) \left( 2l^{2}\omega
b_{1}-2b_{3}-l^{2}\omega ^{2}\right) \right] \sin (4H)-\frac{2l^{2}X_{2}}{%
\lambda }\sin (2H)=0,
\end{split}%
\end{equation}%
where
\begin{equation}
X_{2}(r)=8\lambda \left( l^{2}b_{1}(\omega
-2b_{1})+2b_{2}^{2}+b_{3}(1+2b_{3})\right) \ .
\end{equation}

The closeness with the situation in which one has (instead of the dynamical
Maxwell field) a non-vanishing chemical potential is useful in this case as
well. Indeed, by requiring
\begin{equation}
X_{2}=\lambda \left( l^{2}\omega ^{2}-1\right) =\text{const.} ,
\label{condin3}
\end{equation}
and
\begin{equation}
b_{3}(r)=l^{2}\omega b_{1}(r)-\frac{l^{2}\omega ^{2}}{4}-\frac{1}{4}\ ,
\label{condin4}
\end{equation}
not only the equation for the time-crystal profile becomes solvable (as it
is reduced to a quadrature) but also the full Maxwell equations reduce
consistently to a scalar ODE for $b_{1}(r)$.

All in all, using the ansatz in Eqs. \eqref{EMpotans1}, (\ref{ansnew}), (\ref%
{condin3}) and (\ref{condin4}) (in the line element in Eq. (\ref{metric3})
with coordinates (\ref{cotc})) the \textit{full coupled Skyrme Maxwell system%
} made by Eqs. (\ref{nonlinearsigma1}) and (\ref{maxwellskyrme1}) in a
topologically non-trivial sector (as it will be shown below) \textit{can be
reduced consistently to the following solvable system of two coupled ODEs
for }$H(r)$ and $b_{1}(r)$ (a detailed derivation of this result can be encountered in the appendix)
\begin{equation}
b_{1}^{\prime \prime }-\frac{K}{8}(\omega -4b_{1})\sin ^{2}(H)\left(
l^{2}(\lambda \omega ^{2}-8)-\lambda +\lambda \left( l^{2}\omega
^{2}-1\right) \cos (2H)-8\lambda H^{\prime 2}\right) =0\ ,  \label{b1redtc}
\end{equation}%
\begin{equation}
2\left( \lambda \left( l^{2}\omega ^{2}-1\right) \cos ^{2}(H)-4l^{2}\right)
H^{\prime \prime }+\left( l^{2}\omega ^{2}-1\right) \sin (2H)\left(
l^{2}-\lambda H^{\prime 2}\right) =0\ .  \label{profredtc}
\end{equation}%
Hence, also in this case the recipe is to determine the profile $H(r)$ (as
the corresponding equation is solvable) and then to replace the result into
the equation for $b_{1}(r)$ which becomes a simple linear non-homogeneous
equation in which there is an effective potential which depends on $H(r)$.
The other components of the gauge potential are determined solving the
simple algebraic conditions in Eqs. (\ref{condin3}) and (\ref{condin4}). The
above system allows to clearly disclose many features of the gauged
time-crystals (and, more in general, of the $U(1)$ gauged Skyrme model)
which are close to a ``dual superconductivity".

The first integral of \eqref{profredtc} which allows to reduce it to quadratures is
\begin{equation}
  \left(4 l^2+\lambda \left( 1-l^2  \omega^2\right) \cos ^2(H)\right) H^{\prime 2}  -\frac{1}{2} l^2 \left(1-l^2 \omega ^2\right) \cos (2 H) = I_0,
\end{equation}
where $I_0$ is determined by the boundary conditions. Once $H(r)$ is known, Eq. \eqref{b1redtc} can be analyzed with the standard tools of the theory of linear ordinary differential equations.

As we did in the previous section for the Skyrmion, we also calculate here
for the time crystal, the non vanishing winding number that is produced by
\begin{equation}
\begin{split}
B_{2}& =\frac{\epsilon _{2ijk}}{24\pi ^{2}}\Big[\mathrm{Tr}\left(
R_{i}R_{j}R_{k}\right) -3\partial _{i}\left( A_{j}\mathrm{Tr}\left( \tau
_{3}\left( U^{-1}\partial _{k}U+\partial _{k}UU^{-1}\right) \right) \right) %
\Big] \\
& =\frac{\omega }{8\pi ^{2}}\sin (2H)H^{\prime }+\frac{1}{4\pi ^{2}}\partial
_{r}\left( \cos ^{2}(H)\left( b_{1}-\omega b_{3}\right) \right) ,
\end{split}%
\end{equation}%
where the Latin indices of the previous relation assume the values $0,1,3$
and the resulting integral is

\begin{equation}
W=\int B_{2}drd\left( \omega \gamma \right) d\phi =1+2\Big[\cos
^{2}(H(r))\left( \frac{b_{1}(r)}{\omega }-b_{3}(r)\right) \Big]_{0}^{2\pi
}=1-2\left( \frac{b_{1}(0)}{\omega }-b_{3}(0)\right),   \label{windingtc}
\end{equation}%
if we consider $r\in \lbrack 0,2\pi ]$, $\omega \gamma \in \lbrack 0,4\pi ]$%
, $\phi \in \lbrack 0,2\pi ]$ and $H(2\pi )=\pi /2$, $H(0)=0$.

However, a \textquotedblleft normal\textquotedblright \thinspace\ topological
charge is also present here due to the correction from the
electromagnetic potential. By taking $B_{0}$ as defined in \eqref{tcharge}
as an integral over spatial slices we obtain %\begin{equation}
\begin{equation*}
B=\int B_{0}drdzd\phi =-2\Big[\cos ^{2}(H(r))b_{2}(r)\Big]_{0}^{2\pi
}=2b_{2}(0),
\end{equation*}%
%
%\end{equation}%
with the same boundary values used as in \eqref{windingtc} with the
difference now that we have $z$ in place of $\gamma $ for which we consider $%
z\in \lbrack 0,2\pi ]$. The charge $B$ is non zero as long as $b_{2}(0)\neq 0
$.
\begin{figure}[h]
\centering
\hspace{8mm} \subfloat[][Energy density $T_{00}$ of the Time Crystal]{\includegraphics[width=.40 \textwidth]{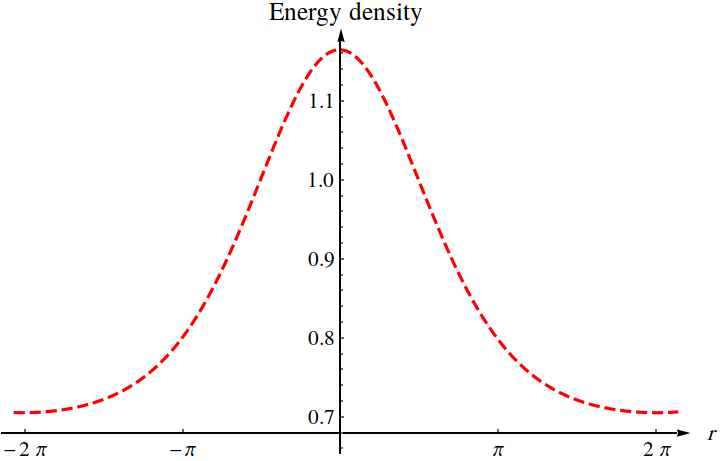}}
\hspace{0mm}
\subfloat[][Time Crystal profile]{
	\includegraphics[width=.41 \textwidth]{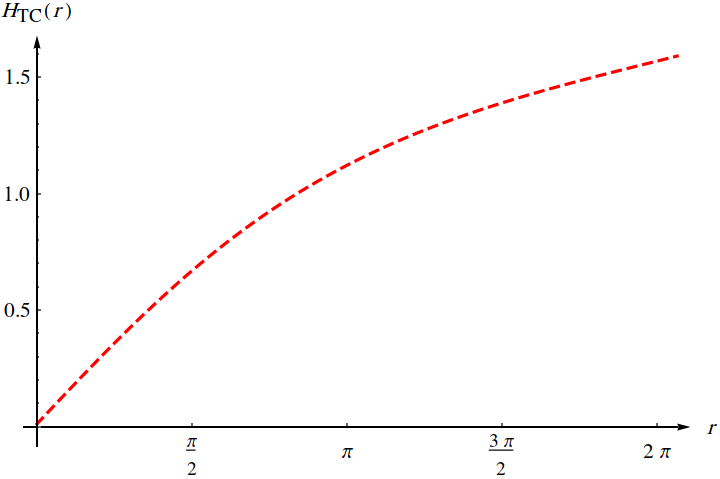}}
\hspace{0mm}
\subfloat[][Gauge potential $A_\mu$]{\includegraphics[width=.40\textwidth]{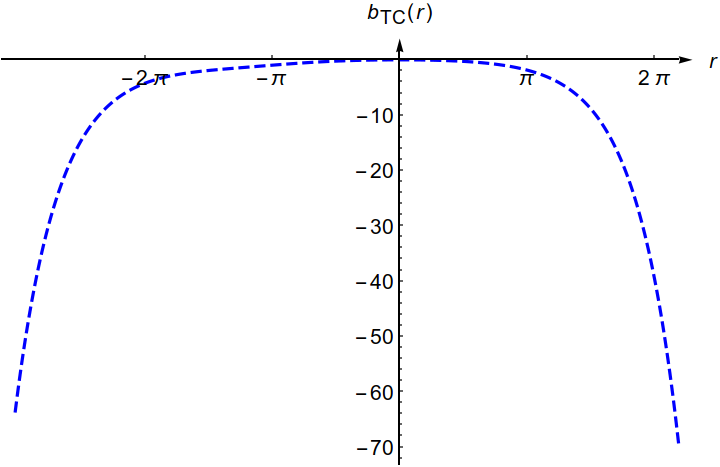}
}
\hspace{0mm}
\subfloat[][Electric field $E\equiv b'(r)$]{\includegraphics[width=.40 \textwidth]{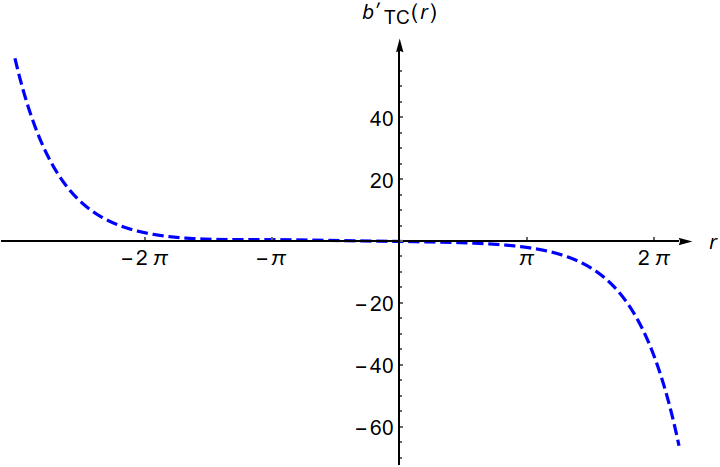} }
\caption{The solutions for the Eqs.\eqref{b1redtc} and \eqref{profredtc} correspond to the values: $\lambda=0.04$, $l=0.47$, $\omega=0.95$, $K=1.00$, $p=1.00$, and $q=1.00$. Solving for $b_{TC}\equiv b_1 $. The above plots show clearly the suppression of the electric field in the core of the time crystal.}  \label{Fig2}
\end{figure}

\section{Extended duality}

In this section we show that an extended electromagnetic duality exists
between the gauged Skyrmion and the gauged time-crystal constructed above.
This means that, in order to disclose such duality, one not only needs to
interchange electric and magnetic components in a suitable way, but also to transform certain parameters
of the gauged solitons.

In other words, the question we want to answer in this section is: how do the usual duality
transformations of the electromagnetic field have to be generalized so as to act on
the Skyrmions and time-crystals considered here in such a way that the field
equations Eqs. (\ref{fullm1}) and (\ref{fullsk1}) (corresponding to the
gauged Skyrmion) are mapped into the field equations Eqs. (\ref{b1redtc})
and (\ref{profredtc}) (corresponding to the gauged time-crystal)?

Let us take\footnote{%
We will analyze here only how to extended electromagnetic duality in the
integrable sectors considered above. However, we hope to come back on the
intriguing issue on how to extended duality to more general configurations
of the $U(1)$ gauged Skyrme model in a future publication. } the simplest
non-trivial cases of gauged configurations we examined above. For the
Skyrmion we have the profile equation \eqref{fullsk1}, which for $p=q=1$
reduces to
\begin{equation}
2\left( 2l^{2}+\lambda \cos ^{2}(H)\right) H^{\prime \prime }+\sin
(2H)\left( l^{2}-\lambda H^{\prime 2}\right) =0,  \label{fullsk11}
\end{equation}%
while the for the electromagnetic potential components we get - from
relations \eqref{condin1} and \eqref{condin2} -
\begin{equation}
b_{1}=\pm \frac{1-4b_{3}}{2\sqrt{2}l},\quad b_{2}=-b_{3},  \label{dualb21}
\end{equation}%
with $b_{3}$ being given by the differential equation
\begin{equation}
b_{3}^{\prime \prime }+\frac{K}{2}(1-4b_{3})\sin ^{2}(H)\left(
2l^{2}+2\lambda H^{\prime 2}+\lambda \cos ^{2}(H)\right) =0\ .
\label{fullm11}
\end{equation}

Let us now consider the corresponding time crystal equations, where - in
order to avoid confusion - we denote the potential as $A_{\mu
}=(a_{1}(r),0,a_{2}(r),a_{3}(r))$ (namely, we label differently the
components). The profile equation is, of course, given by \eqref{profredtc}
with the potential components related as
\begin{equation}
a_{2}=\pm \frac{1}{4}l\sqrt{1-l^{2}\omega ^{2}}(\omega -4a_{1}),\quad
a_{3}=l^{2}\omega a_{1}-\frac{l^{2}\omega ^{2}}{4}-\frac{1}{4},
\label{duala23}
\end{equation}%
and with $a_{1}$ determined by the following equation
\begin{equation}
a_{1}^{\prime \prime }-\frac{K}{8}(\omega -4a_{1})\sin ^{2}(H)\left(
l^{2}(\lambda \omega ^{2}-8)-\lambda +\lambda \left( l^{2}\omega
^{2}-1\right) \cos (2H)-8\lambda H^{\prime 2}\right) =0\ .  \label{duala1}
\end{equation}%
These are just equations \eqref{condin3}, \eqref{condin4} and \eqref{b1redtc}
with the new labeling of the components.

An immediate observation is that profile equations \eqref{profredtc} and %
\eqref{fullsk11} become identical if we set $\omega =-\frac{i}{l}$. Then, it
is an easy task to see that \eqref{duala1} and \eqref{duala23} are mapped to %
\eqref{fullm11} and \eqref{dualb21} under the linear transformation
\begin{equation}
a_{1}=\frac{i}{l}b_{2},\quad a_{2}=i\,l\,b_{1},\quad a_{3}=-b_{3}.
\label{dualtr}
\end{equation}%
The appearance of the imaginary units is not alarming, since one also needs
a suitable imaginary scaling in the relevant coordinates to map one spacetime
metric to the other. Notice that the imaginary part of the transformation
involves only the $\gamma $ and $z$ components of $A_{\mu }$. Hence, the end
result after performing such a transformation is a real electromagnetic tensor of
the Skyrmion case.

We have to note, however, that transformation \eqref{dualtr} is not unique.
There are other linear transformations that map the two set of equations to
each other by mixing the electric and magnetic components. However, %
\eqref{dualtr} belongs to a smaller class of transformations that associates
the electric component of the time crystal potential $a_{1}$ with the purely
magnetic components of the Skyrmion, namely $b_{2}$ and $b_{3}$. In
particular, this property is respected by any linear transformation of the
form $a_{i}=L_{ij}b_{j}$ as long as the following set conditions hold
\begin{equation*}
L_{13}=L_{12}-\frac{i}{l},\quad L_{21}=i\,l,\quad L_{23}=L_{22},\quad
L_{11}=L_{31}=0,\quad L_{33}=L_{32}-1.
\end{equation*}%
Of course, the free parameters appearing in the above transformation must be chosen each time
in such a way so that the end result is strictly real. In the following table we can see how the gauged Skyrmion and the time crystal (T.C.) components of the electromagnetic field are mapped into each other as well as $H$, $A$ and $G$ (in relations \eqref{pions2.25} and \eqref{pions2.26}) that are involved in the generalized Hedgehog ansatz. So we can see that the two configurations correspond to  an interchange between the electric and one of the magnetic components that looks like a duality relation as seen in the plane formed by the $x^1$ and $x^3$ components. One can call this transformation an extended or generalized duality.
\begin{equation*}
  \begin{tabular}{|@{\hspace{2em}}l@{\hspace{3em}}|@{\hspace{2em}}l@{\hspace{2em}}|}
    \hline
    \multicolumn{2}{|l|}{\text{gauged Skyrmion} \; $\longrightarrow$ \; \text{gauged T.C.}} \\ \hline
     $E_1$ & $-B_3$  \\ \hline
    $B_2$ & $-B_2$    \\ \hline
    $B_3$ & $E_1$     \\
    \hline
    $(H,A,G)$ & $(H,A,G)$    \\
    \hline
  \end{tabular}
\end{equation*}

It is a very surprising result that a sort of a duality symmetry exists, which maps the gauged Skyrmion into the gauged time crystal. Thus, if such extended duality transformations discussed here would have been
known in advance, one could have found that time-crystals exist just applying such
transformations to the gauged Skyrmion. Moreover, the plots in Fig. \ref{Fig1} and
Fig. \ref{Fig2}  clearly show that, as the magnetic field is suppressed in the gauged
Skyrmion core, the electric field is suppressed in the gauged time-crystal
core. Thus, as gauged Skyrmions have some features in common with
superconductor, gauged time-crystals have some features in common with dual
superconductor. We hope to come back on this very interesting issue and on
its possible relevance in Yang-Mills theories in a future publication.

\section{External periodic fields}

In this section, we will discuss an approximation which can be of
practical importance in many applications from nuclear physics to
astrophysics.

We have been able to construct analytically two different types of
topologically non-trivial configurations of the full (3+1)-dimensional $U(1)$
gauged Skyrme model (which are dual to each other in the electromagnetic
sense). Thus, it is natural to ask why we should analyze approximated
solutions as we have the exact ones.

The obvious reason is that, in this way, we will be able to discuss
electromagnetic fields more general than the ones leading to the exact
solutions discussed in the previous sections. In particular, it is
interesting to discuss the physical effects of time-periodic electromagnetic
fields (which do not belong to the class leading to the above exact
solutions).

Here it will be considered the case in which the Skyrme configuration is
fixed and not affected by the electromagnetic field (as in \cite{Fab1})
which is slowly turned on to get a tiny time-periodic electromagnetic field
in these Skyrme background solutions. In this case, the Skyrme background
plays the role of an effective medium for the Maxwell equations. Very
interesting is the situation in which the background is a time-crystal as
the reaction of the time-dependent Maxwell perturbation to the presence of
the time-crystal critically depends on the ratio between the frequency of
the perturbation and the frequency of the time-crystal.

\subsection{Tiny time periodic fields in Skyrme background}

Let us consider the approximate situation where we introduce a small enough
electromagnetic field, so as to not consider its effect on the profile
equations. Additionally, we demand that it is periodic in time in one of its
components
\begin{equation}
A_{\mu }=\left( b_{1}(r),0,b_{2}(r)\cos (\Omega \gamma ),b_{3}(r)\right) .
\end{equation}%
The charge is conserved $\partial _{\mu }J^{\mu }=0$, while the Maxwell
equations constitute a compatible system of differential equations. The one
that corresponds to $b_{2}$ is
\begin{equation}
\frac{b_{2}^{\prime \prime }}{b_{2}}=\frac{K}{2}\left[ \lambda \left(
8H^{\prime 2}+2\left( 1-l^{2}\omega ^{2}\right) \cos ^{2}(H)\right) +8l^{2}%
\right] \sin ^{2}(H)-l^{2}\Omega ^{2},  \label{eqb2}
\end{equation}%
and by considering the approximation $b_{2}<<1$ we are led to the single
profile equation
\begin{equation}
H^{\prime \prime }=\frac{l^{2}\lambda \omega ^{2}\sin (4H)}{4\left(
l^{2}\left( \lambda \omega ^{2}-4\right) -\lambda \right) },
\label{profilenoem}
\end{equation}
with the corresponding constant of motion
\begin{equation}
\frac{l^{2}\lambda \omega ^{2}\cos (4H)}{16\left( l^{2}\left( \lambda \omega
^{2}-4\right) -\lambda \right) }+\frac{1}{2}H^{\prime 2}=I_{0}.
\end{equation}%
With the help of the latter, we can express \eqref{eqb2} as
\begin{equation}
\frac{b_{2}^{\prime \prime }}{b_{2}}=\frac{K}{2}\left( x^{2}-1\right) \left(
\frac{2l^{4}\left( \lambda \omega ^{2}-4\right) \left( \lambda x^{2}\omega
^{2}-4\right) +l^{2}\lambda \left( \lambda \left( 8x^{4}-10x^{2}+1\right)
\omega ^{2}+8\right) }{l^{2}\left( \lambda \omega ^{2}-4\right) -\lambda }%
-2\lambda \left( 8I_{0}+x^{2}\right) \right) -l^{2}\Omega ^{2},
\label{eqb2periodic}
\end{equation}%
where $x=\cos (H)$. From the form of \eqref{eqb2periodic} we can deduce that
nature of the solution strongly depends on the sign of
the right hand side. If the sign is negative one expects a periodic type of
behaviour. On the other hand, if it is positive, we rather expect an exponential kind of behaviour.

Clearly, the appearance of this two possibilities has to do with the value
of the frequency $\Omega $ of the field and its relation to the rest of the
parameters of the model.

In general one can consider the function
\begin{equation}
f(x)=\left( x^{2}-1\right) \left( \frac{2l^{4}\left( \lambda \omega
^{2}-4\right) \left( \lambda x^{2}\omega ^{2}-4\right) +l^{2}\lambda \left(
\lambda \left( 8x^{4}-10x^{2}+1\right) \omega ^{2}+8\right) }{l^{2}\left(
\lambda \omega ^{2}-4\right) -\lambda }-2\lambda \left( 8I_{0}+x^{2}\right)
\right),
\end{equation}%
which at most possesses five extrema. The value $x=0$ is always a global
extremum, for the rest of the values of $x$ in $[-1,1]$ one may have from
none up to four extrema depending on the parameters $\lambda $, $l$, $\omega
$ and $I_{0}$.

For example when $l=\omega =\lambda =1$, $I_{0}=-1/2$ one gets five extrema
in the region $x\in \lbrack -1,1]$, of which, $x=0$ is a global maximum; on
the other hand, when $l=\omega =\lambda =1$, $I_{0}=-1$ one gets only one
extremum, $x=0$, which now is a minimum.

In any case, it is possible to arrange the external field frequency $\Omega $
so that the right hand side of \eqref{eqb2periodic} has a clearly positive
or negative sign. The critical value for this is $\Omega _{cr}=\frac{K}{%
2l^{2}}f(0)$, where
\begin{equation}
f(0)=\frac{\lambda \left( 4l^{2}+\lambda \right) }{l^{2}\left( 4-\lambda
\omega ^{2}\right) +\lambda }+8l^{2}+\lambda \left( 16I_{0}-1\right).
\end{equation}%
If $f(0)$ is a maximum, we need to have $\Omega >\Omega _{cr}$ in order to
obtain a periodic type of behaviour. Alternatively if $f(0)$ is a minimum, the condition $%
\Omega <\Omega _{cr}$ leads to an exponential type behaviour.

This simple analysis shows that the reaction of a time periodic Maxwell
perturbation to the presence of a time-crystal strongly depends on the
relations between the frequency of the Maxwell perturbation and the
parameters characterizing the time-crystal.

\section{Conclusions and perspectives}

\label{conclusions}

Using the generalized hedgehog approach we have constructed the first
analytic and topologically non-trivial solutions of the $U(1)$ gauged Skyrme
model in (3+1)-dimensional flat spacetimes at finite volume. There are two types of gauged solitons. Firstly, gauged Skyrmions
living within a finite volume appear as the natural generalization of the
usual Skyrmions living within a finite volume. The second type of gauged
solitons corresponds to gauged time-crystals. These are smooth solutions of
the $U(1)$ gauged Skyrme model whose periodic time-dependence is protected
by a topological conservation law. Interestingly enough, electromagnetic
duality can be extended to include these two types of solitons. Gauged Skyrmions manifest very interesting
similarities with superconductors while gauged time-crystals with dual
superconductors.

Due to the relations of the Skyrme model with low energy limit of QCD, the
present results can be useful in many situations in which the back reaction
of baryons on Maxwell field (and \textit{viceversa}) cannot be neglected
(this is especially true in plasma physics and astrophysics).

It is a very interesting issue (on which we hope to come back in a future
publication) to understand the relevance of the present results in
Yang-Mills theory. From the technical point of view, the tools
which allowed the construction of the present gauged configurations have
been extended to the Yang-Mills case as well (see \cite{canYM1}, \cite%
{canYM2}, \cite{canYM3} and references therein). Thus, it is natural to
wonder whether time-crystal can be defined in Yang-Mills case as well. The
present analysis suggests that this construction could shed some light on
the dual superconductor picture.

\begin{acknowledgements}
This work has been funded by the Fondecyt grants 1160137, 1161150, 3150016
and 3160121. The Centro de Estudios Cient\'{\i}ficos (CECs) is funded by the
Chilean Government through the Centers of Excellence Base Financing Program
of Conicyt. DH and LA are partially founded by Conicyt grant 21160649 and 21160827, respectively.
\end{acknowledgements}

\appendix

\section{Derivation of the reduced system of equations \eqref{Maxgen}, \eqref{skyrmeeq1} and \eqref{tcprof}}

In this section we demonstrate how equations \eqref{Maxgen}, \eqref{skyrmeeq1} and \eqref{tcprof} are obtained from the general field equations \eqref{nonlinearsigma1} and \eqref{maxwellskyrme1}; thanks to the generalized Hedgehog ansatz \cite{canfora3,canfora4,canfora4.5,yang1}, which remarkably enough still holds when the Skyrme field is coupled to Maxwell theory.

It can be easily seen that, under the choice \eqref{ans1} for the Euler angles and \eqref{EMpotans1} for the electromagnetic potential, the three components of $R_\mu=R^i_\mu t_i$, $i=1,2,3$ in Eq. \eqref{sky2.5} read (the order of the spacetime coordinates in the gauged Skyrmion case is $x^\mu = (z,r,\gamma,\phi)$)
\begin{subequations} \label{Rmuexpr1}
\begin{align}
  R^1_\mu &= \left(b_1 \cos (q \phi ) \sin (2 H) , -\sin (q \phi )H', \left(\frac{p}{2}+b_2\right) \cos (q \phi ) \sin (2 H),b_3 \cos (q \phi ) \sin (2 H) \right),  \\
  R^2_\mu &= \left( b_1 \sin (q \phi )\sin (2 H) , \cos (q \phi ) H', \left(\frac{p}{2}+b_2\right) \sin (q \phi ) \sin (2 H), b_3 \sin (q \phi ) \sin (2 H) \right), \\
  R^3_\mu &= \left(-2 b_1 \sin ^2(H),0,\frac{p}{2} \cos (2 H)-2 b_2 \sin ^2(H),\frac{q}{2}-2 b_3 \sin ^2(H)\right) .
\end{align}
\end{subequations}
With the help of the latter, the electromagnetic current vector can be computed through \eqref{current} to be
\begin{equation} \label{Jmuexpr1}
  J^\mu = \frac{K}{2l^2}\left(M_{1I}b_I + N_1,0,M_{2I}b_J + N_2, M_{3I}b_J + N_3 \right), \quad I=1,2,3
\end{equation}
with the expressions $M_{IJ}$ and $N_J$ being given by \eqref{Mmatrix1} and \eqref{Nvector1} respectively. It can be easily verified that $\nabla_\mu J^\mu =0$ holds as an identity for the previous expression.

By using \eqref{Rmuexpr1}, in the three gauged Skyrme equations
\begin{equation} \label{threecompprof}
D^{\mu }\left( R^i_{\mu }t_i+\frac{\lambda }{4}\left[ R^{\nu },G_{\mu \nu }\right]^i t_i
\right) =: E^i t_i =0,
\end{equation}
the latter become
\begin{subequations} \label{deceq}
\begin{align}
  E^1 & = -\frac{\sin (q \phi )}{16 l^4} \mathcal{A}(r)=0, \\
  E^2 & =  \frac{\cos (q \phi )}{16 l^4} \mathcal{A}(r)=0, \\
  E^3 & \equiv 0 ,
\end{align}
\end{subequations}
where
\begin{equation}
  \begin{split}
    \mathcal{A}(r) =& 4 \left(8 \lambda  \sin ^2(H) \left(-2 l^2 b_1^2+ b_2 (2 b_2+p)-b_3(q -2 b_3)\right)+4 l^2+\lambda  \left(p^2+q^2\right)\right) H'' \\
    & -16 \lambda  \sin (2 H) \left(2 l^2 b_1^2-b_2 (2 b_2+p)+b_3 (q-2 b_3)\right) (H^{\prime})^2 \\
    & -32 \lambda  \sin ^2(H) \left(4 l^2 b_1 b_1'-(4 b_2+p) b_2'+(q-4 b_3) b_3'\right) H' \\
    & + \lambda   \left(4 l^2 b_1^2 \left(p^2+q^2\right)-(2 q b_2+p (q-2 b_3))^2\right) \sin (4 H) \\
    & + 16 l^2  \left(2 l^2 b_1^2-b_2 (2 b_2+p)+ b_3 (q-2 b_3)\right) \sin (2 H)
  \end{split}
\end{equation}
We can see that the $t_3$ component becomes identically zero, while the other two are proportional after the substitution of all the involved quantities. The remaining $\phi$ variable is decoupled from $r$ and the system is reduced to the single equation, $\mathcal{A}=0$, for $H(r)$, which we have expressed in a more compact form in \eqref{skyrmeeq1}.  At the same time, the current $J^\mu$, as given by \eqref{Jmuexpr1}, is only $r$ dependent and leads to the Maxwell set of equations \eqref{Maxgen}.

The exact same thing can be repeated for the profile equation of the gauged time crystal \eqref{tcprof}. This time we have to consider \eqref{eultc} for the Euler angles, with the help of which the three $R_\mu$ components are written as (remember that now $x^\mu = (\gamma,r,z,\phi)$)
\begin{subequations} \label{Rmuexpr2}
\begin{align}
  R^1_\mu &= \left(b_1 \cos (\omega \gamma) \sin (2 H), -\sin (\omega \gamma) H', b_2 \cos ( \omega \gamma  ) \sin (2 H), \left(\frac{1}{2}+b_3\right) \cos(\omega\gamma)\sin(2H)\right),  \\
  R^2_\mu &= \left(b_1 \sin (\omega \gamma) \sin (2 H), \cos (\omega \gamma) H', b_2 \sin ( \omega \gamma  ) \sin (2 H), \left(\frac{1}{2}+b_3\right) \sin(\omega\gamma)\sin(2H)\right), \\
  R^3_\mu &= \left(\frac{\omega }{2}-2 b_1 \sin ^2(H), 0,-2 b_2 \sin ^2(H), \frac{1}{2} \cos (2 H)-2 b_3 \sin ^2(H)\right) .
\end{align}
\end{subequations}
In the same manner the variables are decoupled in the three profile equations \eqref{threecompprof} and the system once more  reduces to the single equation. The $t_3$ component is identically zero, while the rest two are proportional to each other leading to a single equation for $H(r)$, which is given by \eqref{tcprof}. In particular, we obtain
\begin{subequations} \label{deceq2}
\begin{align}
  E^1 & = -\frac{\sin (\omega \gamma )}{16 l^4} \mathcal{B}(r)=0, \\
  E^2 & =  \frac{\cos (\omega \gamma )}{16 l^4} \mathcal{B}(r)=0, \\
  E^3 & \equiv 0 ,
\end{align}
\end{subequations}
with
\begin{equation}
  \begin{split}
  \mathcal{B} = & 4 \left(8 \lambda  \sin ^2(H) \left(l^2 b_1 (\omega -2 b_1)+2 b_2^2+ b_3(1+2 b_3)\right)+l^2 \left(4-\lambda  \omega ^2\right)+\lambda \right) H^{\prime\prime} \\
  & + 16 \lambda  \sin (2 H) \left(l^2 b_1 (\omega -2 b_1)+2 b_2^2+ b_3(1+ 2 b_3) \right) (H^{\prime})^2 \\
  & + 32 \lambda  \sin ^2(H) \left(l^2 (\omega -4 b_1) b_1'+4 b_2 b_2'+(4 b_3+1) b_3'\right) H^{\prime} \\
  & +\lambda   \left( l^2 \omega ^2 + 4 l^2 b_1 (b_1-\omega )  + 4 b_2^2 \left(l^2 \omega ^2-1\right) + 4 l^2 \omega  b_3 (-2 b_1+\omega  b_3+\omega )\right) \sin (4 H) \\
  & -16 l^2 \left(l^2 b_1 (\omega -2 b_1)+2 b_2^2+2 b_3^2+ b_3\right) \sin (2 H).
  \end{split}
\end{equation}
It is easy to verify that $\mathcal{B}=0$ is equivalent to \eqref{tcprof}. Of course the same considerations are also true for the Maxwell equations and relation \eqref{Jmuexpr1} still holds for the current, where now the $M_{IJ}$ and $N_I$ are given by the expressions \eqref{Mmatrix2} and \eqref{Nvector2} respectively.

\end{document}